\documentclass{appolb}
\usepackage{graphicx}
\usepackage{bm}
\usepackage{amssymb,amsmath,latexsym}
\usepackage{xcolor}
\definecolor{darkblue}{RGB}{0,0,196}
\definecolor{darkred}{RGB}{196,0,0}
\usepackage[colorlinks=true,linktocpage=true,linkcolor=darkblue,citecolor=darkblue,urlcolor=darkblue]{hyperref}

\newcommand{\p}{\mathbf{p}}

\def\be{\begin{equation}}
\def\ee{\end{equation}}
\def\ba{\begin{eqnarray}}
\def\ea{\end{eqnarray}}

\begin{document}
\title{Non-equilibrium attractor in high-temperature QCD plasmas%
\thanks{Presented at Quark Matter 2022}%
}
\author{Dekrayat Almaalol\footnote{Speaker.}
\address{University of Illinois at Urbana-Champaign, Urbana, IL 61801, United States}
\\[3mm]
{Kirill Boguslavski % of different affiliation
\address{{Institute for Theoretical Physics, TU Wien, 
1040 Vienna, Austria}}
}
\\[3mm]
{Aleksi Kurkela % of different affiliation
\address{University of Stavanger, 4036 Stavanger, Norway}
}
\\[3mm]
Michael Strickland
\address{Department of Physics, Kent State University, Kent, OH 44242, United States}
}
\maketitle
\begin{abstract}
We establish the existence of a far-from-equilibrium attractor in weakly-coupled gauge theory undergoing 0+1d Bjorken expansion which goes beyond the energy-momentum tensor to the detailed form of the one-particle distribution function. We then demonsrate that the dynamics can be rescaled at intermediate times and represented by universal exponents.  Finally, we assess different procedures for reconstructing the full one-particle distribution function from the energy-momentum tensor along the attractor and discuss implications for the freeze-out procedure used in the phenomenological analysis of ultra-relativistic nuclear collisions
\end{abstract}
  
\section{Introduction}
In the precision era of heavy ion collisions there has been an increased interest in understanding and quantifying the effects of non-equilibrium corrections present at different stages in heavy ion collisions. Non-equilibrium corrections are significantly present and particularly important in two main phases of the dynamical evolution: (I) pre-hydrodynamics (pre-equilibrium), and (II) freeze out (particlization).  Non-equilibrium attractors can be used to explore the approach to equilibrium and have been widely applied to examine the applicability of hydrodynamic theories out-of-equilibrium in many approaches. In this work \cite{PRL125}, we employ a microscopic approach based on quantum chromodynamics (QCD) effective kinetic theory (EKT), which is derived using weak-coupling methods.  The method is applicable at high temperatures and is appropriate for modelling the initial stages of ultrarelativistic heavy-ion collisions. In parametrically isotropic systems,  EKT \cite{Arnold:2002zm,York:2014wja} gives a leading-order accurate description (in $\alpha_s$) of the time evolution of the one-particle distribution function in QCD and allows for a numerical realization of the so-called bottom-up thermalization scenario \cite{Baier:2000sb}.  The description is based on the relativistic Boltzmann equation
\be
-\frac{d f(\p)}{d\tau}+ \frac{p_z}{\tau}\partial_{p_z} f = \mathcal{C}_{1\leftrightarrow 2}[f(\p)] + \mathcal{C}_{2\leftrightarrow 2}[f(\p)] \, ,
\label{eq:ekt1}
\ee
where $f(\p)$ is the gluonic one-particle distribution function (per degree of freedom).  The elastic scattering term $\mathcal{C}_{2\leftrightarrow 2}$ and the effective inelastic term $\mathcal{C}_{1 \leftrightarrow 2}$ include physics of dynamical screening and Landau-Pomeranchuck-Migdal suppression. For the numerical solution of Eq.~\eqref{eq:ekt1}, we discretize $n(\p)= p^2  f(\p)$ on an optimized momentum-space grid and use Monte Carlo sampling to compute the integrals appearing in the elastic and inelastic collisional kernels. The algorithm used is based on Refs.~\cite{York:2014wja,Kurkela:2015qoa,Keegan:2015avk,Arnold:2003zc}.

\section{Non-equilibrium QCD attractor for higher moments}
%--------------------------------------------------
The time evolution of integral moments which characterize the momentum dependence of the distribution function is given by ~\cite{Strickland:2018ayk}
\begin{equation}
{\cal M}^{nm}(\tau) \equiv  \int \frac{d^3 p}{(2\pi)^3} \, p^{n-1} \, p_z^{2m} \, f(\tau,\p) \, ,
\label{eq:genmom1}
\end{equation}
where $p = |{\bf p}|$.  Note that the energy density is given by $\varepsilon=\nu {\cal M}^{20}$, longitudinal pressure  by $P_L = \nu {\cal M}^{01}$, and number density by $n=\nu {\cal M}^{10}$ for $\nu$ degrees of freedom ($\nu = 2 d_A$ for $d_A$ adjoint degrees of freedom). These moments will be scaled by their corresponding equilibrium values with \mbox{$\overline{\cal{M}}^{nm}(\tau) \equiv {\cal M}^{nm}(\tau) / {\cal M}^{nm}_{\rm eq}(\tau)$}, 
where, using a Bose distribution, one obtains
\begin{equation}
{\cal M}^{nm}_{\rm eq} =  \frac{ T^{n+2m+2}\Gamma(n+2m+2) \zeta(n+2m+2)}{2 \pi^2 (2m+1)} \, .
\end{equation}
%
%-----------------
\begin{figure*}[t!]
\includegraphics[width=1\linewidth]{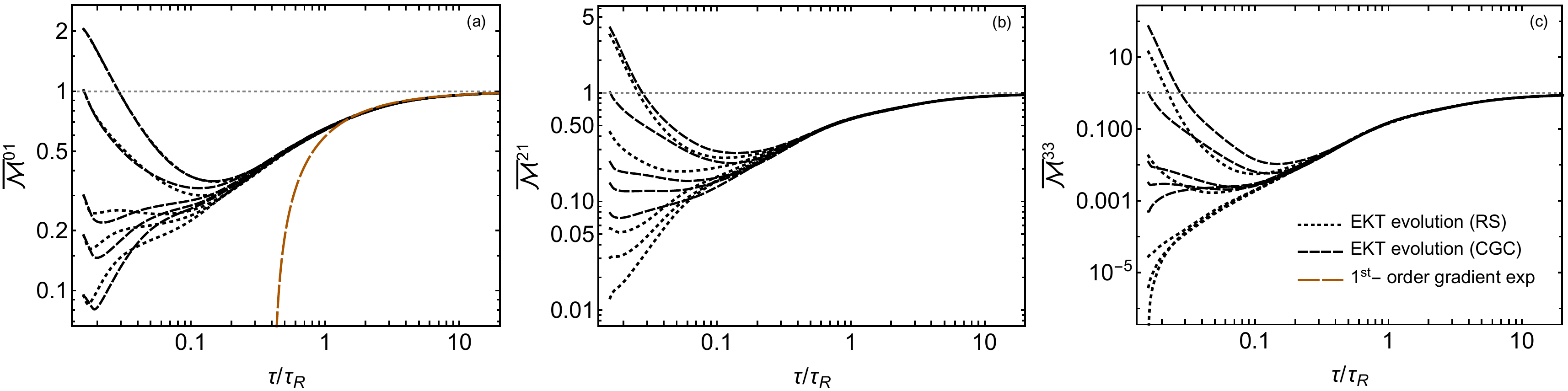}\\
\includegraphics[width=1\linewidth]{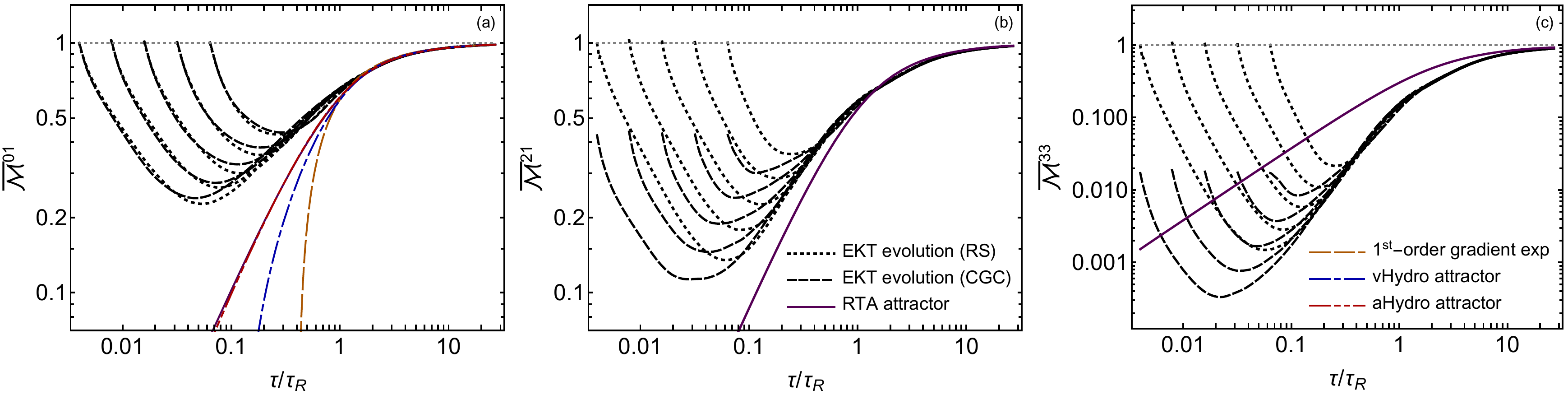}
\caption{Evolution of the scaled moments (a) $\overline{\cal{M}}^{01}$, (b) $\overline{\cal{M}}^{21}$, and (c) $\overline{\cal{M}}^{33}$. The top row corresponds to varying the initial anisotropy ($\xi_0$) and bottom row to varying the initialization time $\tau_0$. Black dotted and dashed lines show the EKT evolution with RS and CGC initial conditions, respectively.  The purple solid line is the exact RTA attractor, the orange long-dashed line is the first-order gradient expansion result, the blue dot-dashed line is the DNMR vHydro attractor, and the red dot-dot-dashed line is the aHydro attractor.} 
\label{fig1}
\end{figure*}
%-------------------------------------
The temperature $T$ here corresponds to the temperature of an equilibrium system with the same energy density, given by 
$T = (30 \varepsilon/\nu \pi^2)^{1/4}$. 
\par
These simulations are initialized with either of the two following initial conditions:  (1) spheroidally-deformed thermal initial conditions which we will refer to as ``RS'' initial conditions \cite{Romatschke:2003ms}
\be
f_{0,{\rm RS}}(\p) = f_{\rm Bose}\!\left(\sqrt{\p^2 + \xi_0 p_z^2 }/\Lambda_0\right) ,
\ee
where $-1 < \xi_0 < \infty$ encodes the initial momentum-anisotropy and $\Lambda_0$ is a temperature-like scale which sets the magnitude of the initial average transverse momentum,  or (2) non-thermal color-glass-condensate (CGC) inspired initial conditions \cite{Kurkela:2015qoa}
\ba
f_{0,\rm CGC}(\p) &=&  \frac{2A}{\lambda} \frac{ \tilde \Lambda_0}{\sqrt{ \p^2 +\xi_0  p_z^2}} e^{-\frac{2}{3}\left(\p ^2 +  \xi_0 \hat{p}_z^2 \right)/\tilde \Lambda_0^2} \, .
\ea
In Fig.~\ref{fig1}, we present results for the evolution of three scaled moments, $\overline{\cal{M}}^{01}$, $\overline{\cal{M}}^{21}$, and $\overline{\cal{M}}^{33}$, in panels (a), (b), and (c), respectively. In the top panel, we fix the initialisation time and examine the existence of the forward attractor or the convergence toward late time equilibrium state of the system. In the bottom panel which corresponds to the ``pullback" attractor or the convergence to the free streaming phase of the dynamics, we vary the initialization time toward asymptotically early times $\tau\sim0$. As we show in Fig.~(\ref{fig1}), different solutions to Eq.~(1) collapse to a universal curve at roughly $\tau/\tau_R \sim 0.5$ which indicates insensitivity to initial anisotropy and occupancy and confirms the existence of an attractor solution \cite{PRL125}.   
\subsection{Re-scaling the turning point}
\begin{figure}[thp]
	\centering
	\includegraphics[width=0.312\textwidth]{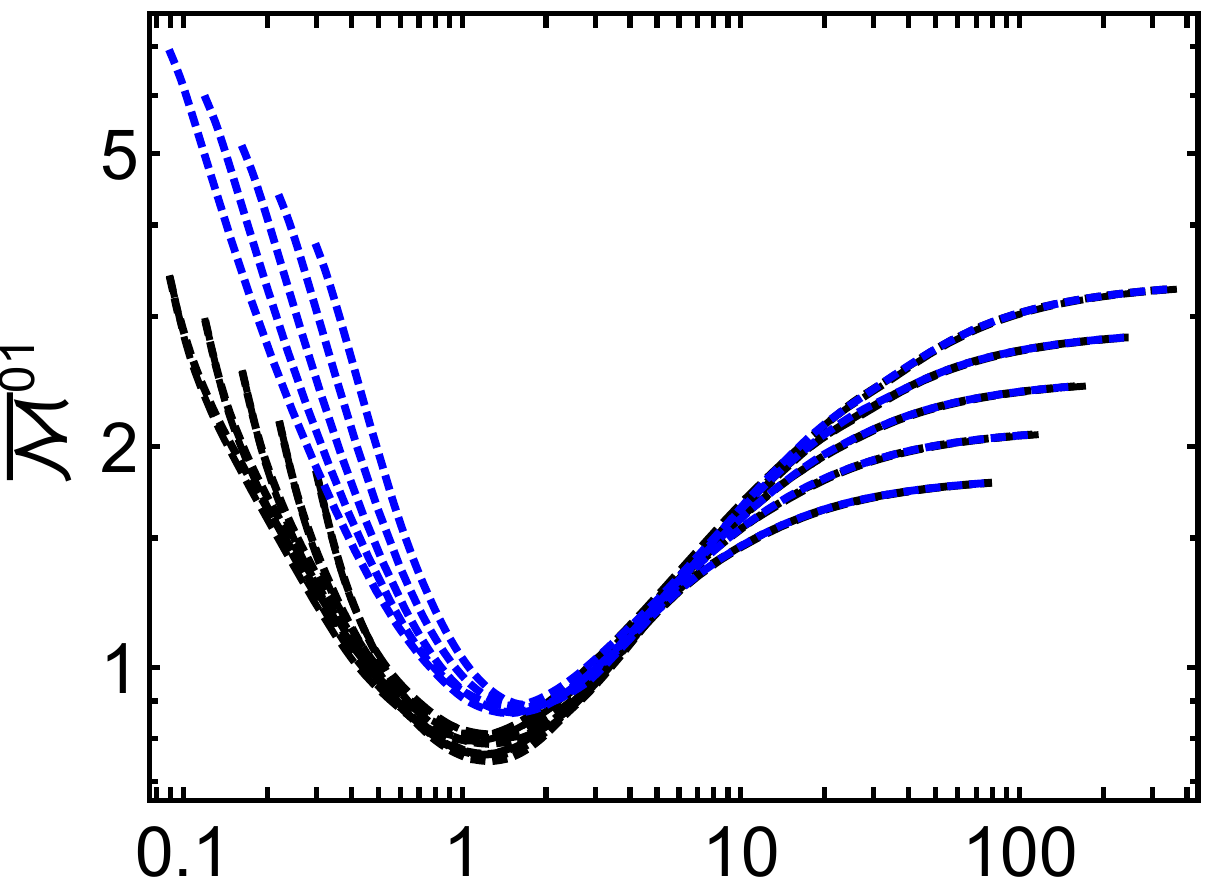}
	\includegraphics[width=0.322\textwidth]{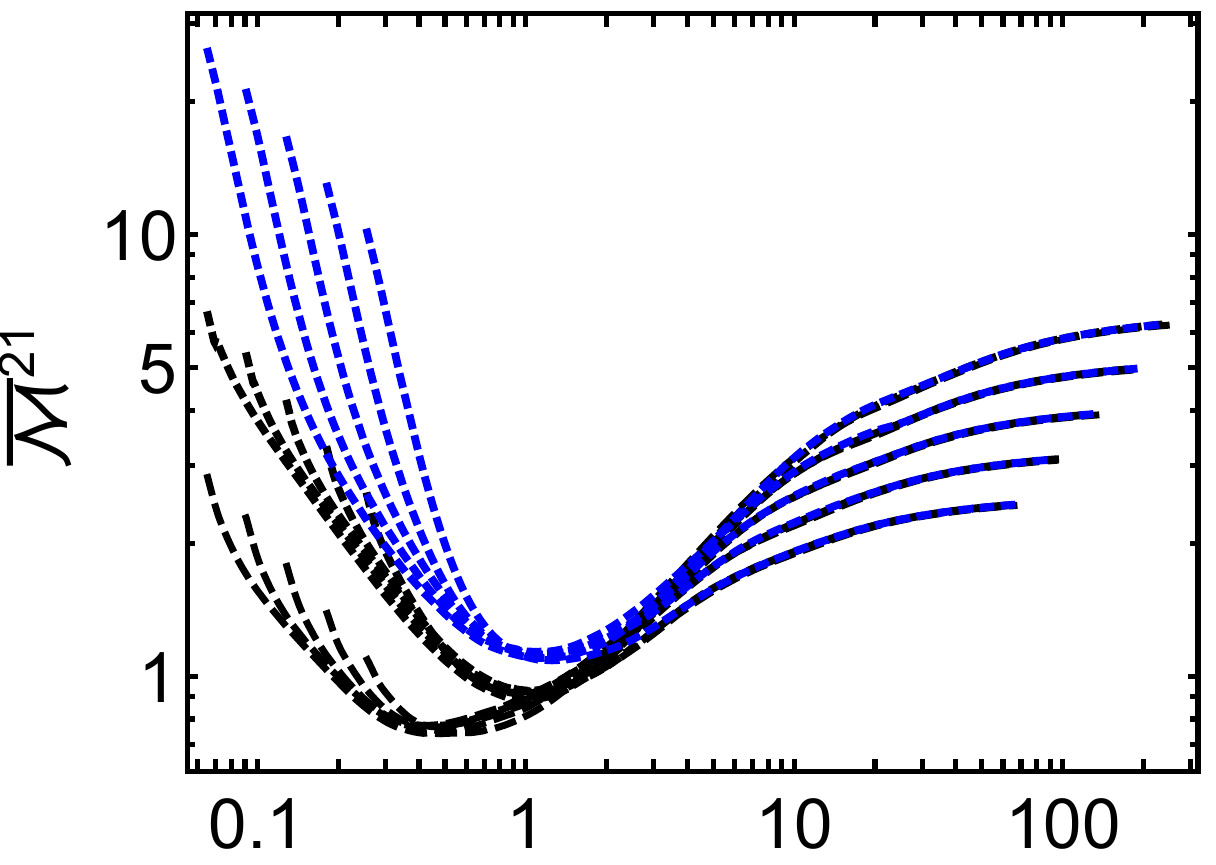}
	\includegraphics[width=0.344\textwidth]{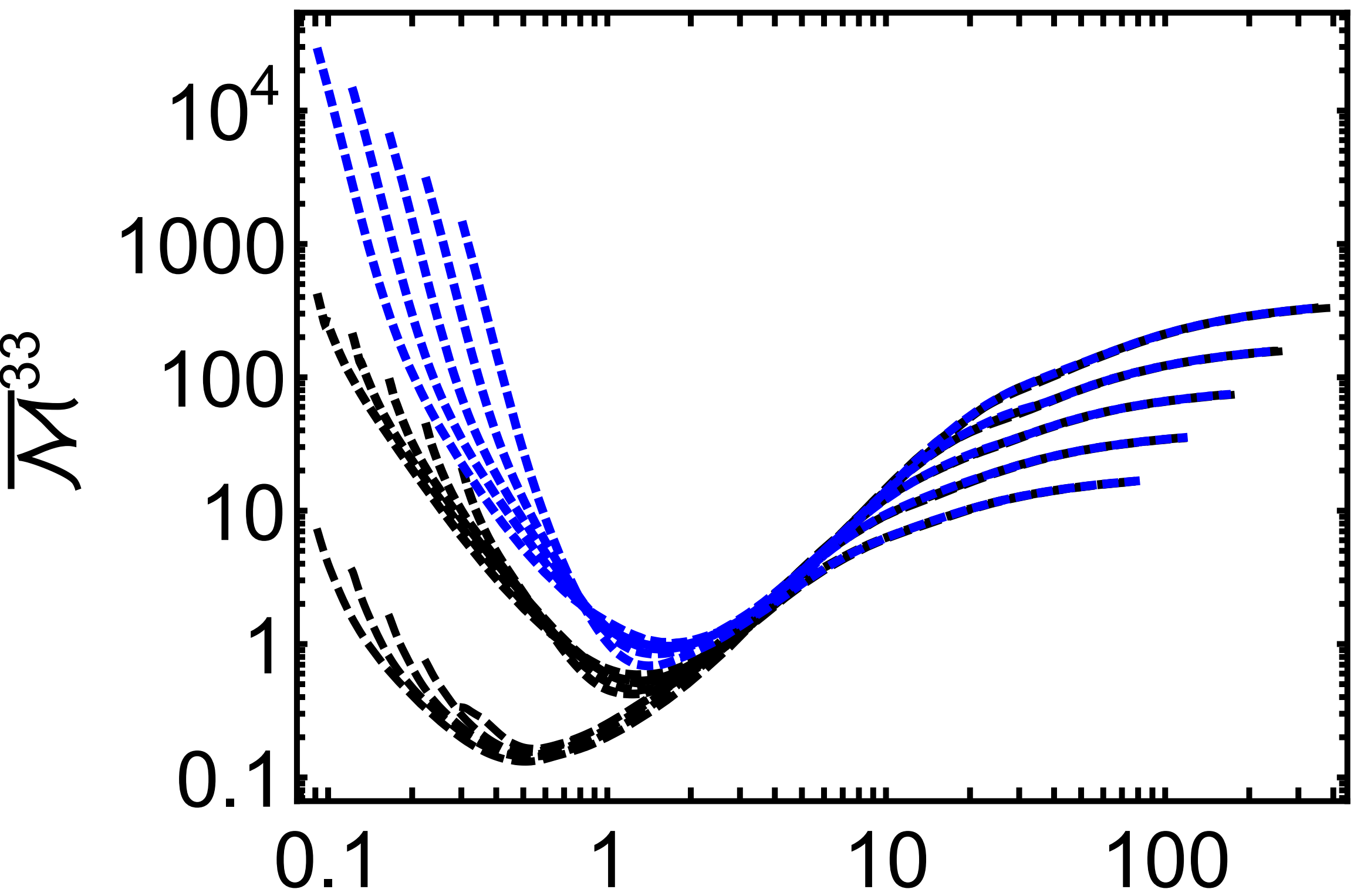}\\
	\includegraphics[width=0.8\textwidth]{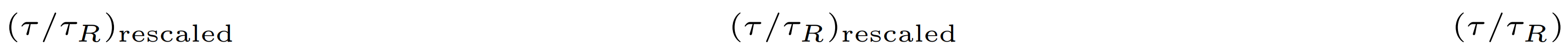}
	\caption{Rescaled moments $\overline{\mathcal M}^{nm} / \omega_0^{A_{nm}}$ as functions of $\omega / \omega_0^{B_{nm}}$ for different sets of initial conditions and initialization times $\omega_0 = \tau_0/\tau_R$. 
 Rescaling a particular moment and the scaled time $\omega$ reproduces a universal rescaling at the turning point at which the microscopic interactions begin to take over the dynamics. }
	\label{fig:midRescaled}
\end{figure}

 %Bottom-up thermalisation scenario is characterized by the competition between the expansion and the interaction rates.%
 In addition to looking for scaling properties at early and late times \cite{PhysRevLett.124.102301,PhysRevD.106.014016}, one can then also investigate whether there is universal scaling dynamics in the initialization time $\omega_0 = \tau_0/\tau_R$ at the turning point.
With $\omega \equiv \tau/\tau_R$, we define the latter as the time $\omega_{nm}(\omega_0)$ of the minimum of the moment $\overline{\mathcal M}^{nm}(\omega_0)$, 
assuming that both $\omega_{nm}$ and $\overline{\mathcal M}^{nm}$ are power laws in $\omega_0$ with $\overline{\mathcal M}^{nm} \sim \omega_0^{A_{nm}}$ and $\omega_{nm} \sim \omega_0^{B_{nm}}$. If we also assume that the scaling exponents for fixed $n,m$ are universal and independent of the initial conditions, we can estimate their value by taking the average of the exponents from the fits to the RS and CGC initial conditions. Then, taking both $A_{nm}$ and $B_{nm}$ to depend linearly on $n$ and $m$, the resulting fits are
\begin{align}
A_{nm} &= - 0.0604726 \, (2 - n) + 0.340507 \, m \, ,\\
B_{nm} &= 0.51845 - 0.0285393 \, n + 0.0452043 \, m\,.
\end{align}
Interestingly, we note that the exponent $B_{nm}$ seems to be approximately $1/2$ with small corrections for different $n,m$ while the exponent $A_{nm}$ shows strong dependence on $m$ and depends on $n$ only weakly \cite{almaalol:2022} 
%as visible in 
(see Fig.~(\ref{fig:midRescaled})). 
%%%%%%%%%%%%%%%%%%%%%%%%%
%%%%%%%%%%%%%%%%%%%%%%%%%%
\section{Reconstructing the one-particle distribution function from $T^{\mu\nu}$}

While the fluid-dynamic theories do not specify the higher moments of the distribution functions, in phenomenological applications it is a common practice to infer the full shape of the distribution from the shear components of the energy-momentum tensor only for use in freeze out. For a given $T^{\mu\nu}$, the linearized viscous correction to the one-particle distribution function, $\delta f$ can be locally computed given an assumption of the collision kernel. Herein, we consider two possible forms for $\delta f$. The (i) \emph{quadratic ansatz}
\be
\frac{\delta f_{(i)}}{f_{\rm eq}(1+f_{\rm eq})} =  \frac{3 \overline\Pi}{16 T^2} ( p^2 - 3 p_z^2 ) \, ,
\label{eq:fa}
\ee
which results from a wide set of models. Here $\overline{\Pi} = \Pi/\epsilon = 1/3 - T^{zz}/\epsilon$. At full leading-order, however, QCD EKT has a more rich structure; for large $p \gg T$, QCD EKT reduces to power law form of the (ii) \emph{Landau-Pomeranchuck-Migdal (LPM) ansatz}
\be
\frac{\delta f_{(ii)}}{f_{\rm eq}(1+f_{\rm eq})} = \frac{16 \overline\Pi}{21 \sqrt{\pi}\, T^{3/2}} \! \left( p^{3/2} - \frac{3 p_z^2}{\sqrt{p}} \right) .
\label{eq:fb}
\ee
This $p^{1.5}$ power-law is numerically close to $\propto p^{1.38}$, which was found to describe the high-momentum region of the full EKT result~\cite{Dusling:2009df}.  Finally, we consider the non-linear (iii) \emph{aHydro freeze-out ansatz} in which one assumes that the distribution function can be approximated by a spheroidally-deformed Bose-distribution $f(p) = f_{\rm Bose}(\sqrt{{\bf p}^2 + \xi p_z^2}/\Lambda)$~\cite{Romatschke:2003ms,Florkowski:2010cf,Martinez:2010sc}. 

%--------------------
\begin{figure*}[t!]
\includegraphics[width=0.99\linewidth]{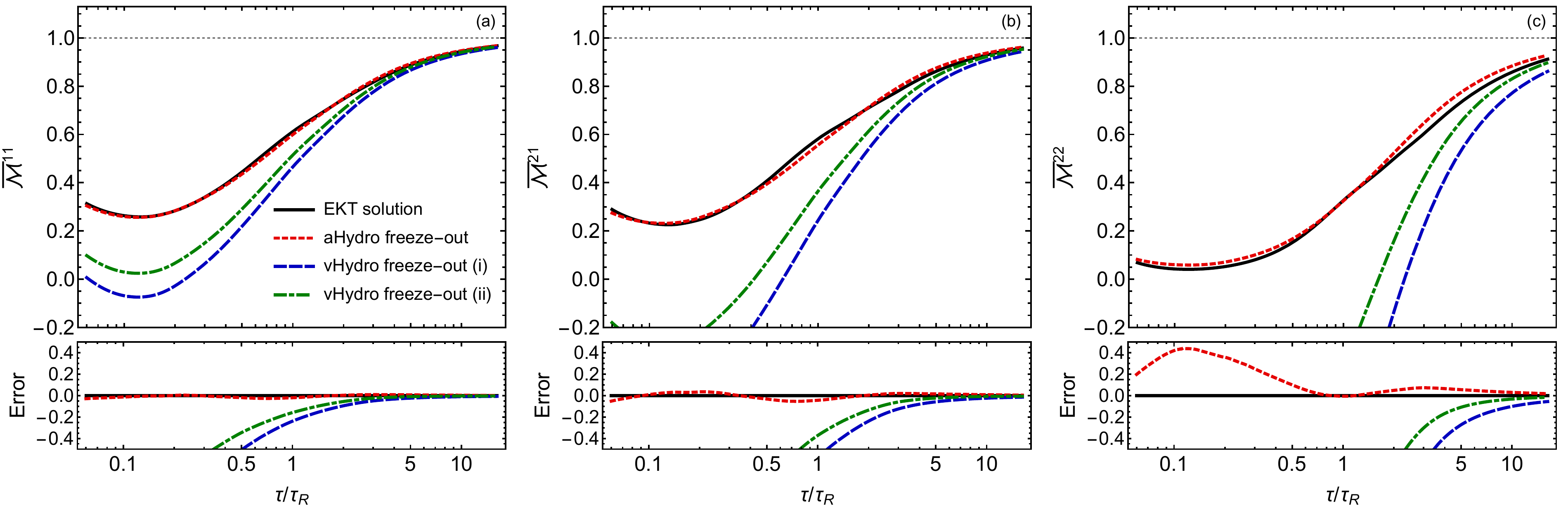}
\caption{Evolution of the scaled moments (a) $\overline{\cal{M}}^{11}$, (b) $\overline{\cal{M}}^{21}$, and (c) $\overline{\cal{M}}^{22}$.  The black solid line is a typical EKT evolution, the red-dashed line is the $P_L$-matched aHydro result for a given moment,  the blue and green dot-dashed lines are the corresponding vHydro results using , respectively.  The relative error shown in the bottom panels is (${\rm approximation}/{\rm EKT} -1$).}
%Eqs.~\eqref{eq:fa} and \eqref{eq:fb}
\label{fig3}
\end{figure*}

The different moments obtained by the above prescriptions are compared to the EKT attractor solution in Fig.~3. At late times $\tau > 5\,\tau_R$, the low-order moments are described within a few percent by all the prescriptions, while some discrepancy remains even at $\tau  \sim 20\,\tau_R$ between the quadratic ansatz (i) and our EKT results. The agreement worsens at earlier times and, around \mbox{$\tau \sim \tau_R$} where the corrections to longitudinal pressure start to become sizable $P_L/P_L^{\rm eq} \sim 65\%$, ${\cal M}^{11}$ exhibits an approximately $20\%$ disagreement between EKT and both linearized ansatze. The disagreement increases for higher moments and at earlier times. In contrast, we observe good agreement between the aHydro ansatz and our EKT results at all times. As a result, when considering higher-moments or applying early-time freeze-out for smaller systems such as peripheral nucleus-nucleus collisions and proton-nucleus collision, the aHydro freeze-out ansatz is favored. 

%\section*{Acknowledgement}
~\\
\textbf{Acknowledgements:} M.S. and D.A. were supported by the U.S. Department of Energy, Office of Science, Office of Nuclear Physics Award No. DE-SC0013470. D.A. is also supported by the US-DOE Nuclear Science Grant No.\ DE-SC0020633. 
K.B.\ is supported in part by the Austrian Science Fund (FWF) project P 34455. 
The authors wish to acknowledge the Ohio Supercomputer Center
project No.\ PGS0251 and 
the Vienna Scientific Cluster (VSC) project 71444 for computational resources.


\begin{thebibliography}{99}

\bibitem{PRL125}
D. Almaalol, A. Kurkela, and M. Strickland,
%``Non-equilibrium attractor in high-temperature QCD plasmas,''
Phys. Rev. Lett. \textbf{125} (2020).
%doi:10.1103/PhysRevLett.125.122302

\bibitem{Arnold:2002zm}
P. Arnold, G. Moore, and L. Yaffe,
%``Effective kinetic theory for high temperature gauge theories,''
JHEP \textbf{01}, 0209353 (2003).
%doi:10.1088/1126-6708/2003/01/030

\bibitem{York:2014wja}
Y. Abraao, C. Mark, A. Kurkela, E. Lu, and G.D. Moore,
%``UV cascade in classical Yang-Mills theory via kinetic theory,''
Physical Review D. \textbf{89}, 074036 (2014).
%doi:10.1103/physrevd.89.074036

\bibitem{Kurkela:2015qoa}
A. Kurkela and Y. Zhu,
%``Isotropization and hydrodynamization in weakly coupled heavy-ion collisions,''
Phys. Rev. Lett. \textbf{115}, 182301 (2015).
%doi:10.1103/PhysRevLett.115.182301

\bibitem{Baier:2000sb}
R. Baier, A.H. Mueller, D. Schiff, and D.T. Son,
%``Bottom up' thermalization in heavy ion collisions,''
Phys. Lett. \textbf{B105}, 0009237 (2001).
%doi:10.1016/S0370-2693(01)00191-5

\bibitem{Keegan:2015avk}
L. Keegan, A. Kurkela, P. Romatschke, W. van der Schee, and Y. Zhu,
%``Weak and strong coupling equilibration in nonabelian gauge theories,''
JHEP \textbf{04}, 031 (2016).
%doi:10.1007/JHEP04(2016)031

\bibitem{Arnold:2003zc}
P. Arnold, G.D. Moore, and L. Yaffe,
%``Transport coefficients in high temperature gauge theories Beyond leading log,''
JHEP \textbf{05}, 051 (2003).
%doi:10.1088/1126-6708/2003/05/051

\bibitem{Strickland:2018ayk}
M. Strickland,
%``The non-equilibrium attractor for kinetic theory in relaxation time approximation,''
JHEP \textbf{12}, 128 (2018).
%doi:10.1007/JHEP12(2018)128

\bibitem{Romatschke:2003ms}
P. Romatschke and M. Strickland,
%``Collective modes of an anisotropic quark gluon plasma,''
Phys. Rev. \textbf{D68}, 036004 (2003).
%doi:10.1103/PhysRevD.68.036004

\bibitem{PhysRevLett.124.102301}
A. Kurkela, W. van der Schee, U.A. Wiedemann, and B. Wu,
%``Early- and Late-Time Behavior of Attractors in Heavy-Ion Collisions,''
Phys. Rev. Lett. \textbf{124} (2020).
%doi:10.1103/PhysRevLett.124.102301

\bibitem{PhysRevD.106.014016}
X. Du, M.P. Heller, S. Schlichting, and V. Svensson,
%``Exponential approach to the hydrodynamic attractor in Yang-Mills kinetic theory,''
Phys. Rev. D \textbf{106} 014016 (2022).
%doi:10.1103/PhysRevD.106.014016

\bibitem{almaalol:2022}
D. Almaalol and K. Boguslavski, forthcoming.

\bibitem{Dusling:2009df}
K. Dusling, G.D. Moore, and D. Teaney,
%``Radiative energy loss and v(2) spectra for viscous hydrodynamics,''
Phys. Rev. \textbf{C81}, 034907 (2010).
%doi:10.1103/PhysRevC.81.034907

\bibitem{Florkowski:2010cf}
W. Florkowski and R. Ryblewski,
%``Highly-anisotropic and strongly-dissipative hydrodynamics for early stages of relativistic heavy-ion collisions,''
Phys. Rev. \textbf{C83}, 034907 (2011).
%doi:10.1103/PhysRevC.83.034907

\bibitem{Martinez:2010sc}
M. Martinez and M. Strickland,
%``Dissipative Dynamics of Highly Anisotropic Systems,''
Nucl. Phys. \textbf{A848} (2010).
%doi:10.1016/j.nuclphysa.2010.08.011

\end{thebibliography}
\end{document}